\input mtexsis
\paper
\singlespaced
\widenspacing
\twelvepoint
\Eurostyletrue
\thicksize=0pt
\sectionminspace=0.1\vsize
%definitions
\def\ga{g_{\sss A}}

\def\nct{\textstyle{{{N}_c}\over 2}}
\def\nc{{N}_c}
\def\yo1{{{f_\pi}^2}}

\def\oneht{\textstyle{1\over 2} }

\def\threeht{\textstyle{3\over 2} }
\def\fiveht{\textstyle{5\over 2} }

\def\sss{\scriptscriptstyle}

%references
\referencelist
\reference{witten}  E.~Witten, \journal Nucl. Phys. B;160,57 (1979)
\endreference
\reference{gervais}  J.L.~Gervais and B.~Sakita,
\journal Phys. Rev. Lett.;52,87 (1984);\hfill\break
\journal Phys. Rev. D;30,1795 (1984);
\endreference
\reference{dashen}  R.~Dashen and A.V.~Manohar,
\journal Phys. Lett. B;315,425 (1993); {\it ibid}, 447 
\endreference
\reference{*dashena}  For a review and extensive references, see
E.~Jenkins, {hep-ph/9803349}
\endreference
\reference{adler} S.L.~Adler, \journal Phys. Rev. Lett.;14,1051 (1965) 
\endreference
\reference{*adlera} W.I.~Weisberger, \journal Phys. Rev. Lett.;14,1047 (1965) 
\endreference
\reference{alg}  S.~Weinberg, \journal Phys. Rev.;177,2604 (1969);\hfill\break
\journal Phys. Rev. Lett.;65,1177 (1990); {\it ibid}, 1181
\endreference
\reference{bron} W.~Broniowski,
\journal Nucl. Phys. A;580,429 (1994), {hep-ph/9402206}
\endreference
\reference{strong}  S.~Weinberg, {\it Strong Interactions at Low Energies},
{hep-ph/9412326}
\endreference
\reference{coleman}  S.~Coleman and E.~Witten, 
\journal Phys. Rev. Lett.;45,100 (1980)
\endreference
\reference{mano} See, for instance, A.V.~Manohar and E.~Jenkins, 
\journal Phys. Lett. B;259,353 (1991) 
\endreference
\reference{alfaro}  R.~de Alfaro, S.~Fubini, G.~Furlan, and G.~Rossetti, 
         {\it Currents in Hadron Physics}, (North-Holland, Amsterdam, 1973)
\endreference
\reference{hemmert} T.R.~Hemmert, B.R.~Holstein and J.~Kambor,
{hep-ph/9712496}
\endreference
\reference{pirjol}  D.~Pirjol and T-M.~Yan,
\journal Phys. Rev. D;57,1449 (1998); {\it ibid}, 5434
\endreference
\reference{beane}  S.R.~Beane and U.~van Kolck, work in progress
\endreference
\endreferencelist
% title page
\titlepage
\obeylines
\hskip4.8in{DOE/ER/40762-160}
\hskip4.8in{U.ofMd.PP\#99-018}
\hskip4.8in{hep-ph/9809328}
\unobeylines
\vskip0.5in
\title
Chiral Multiplets of Large-$\nc$ Ground State Baryons
\endtitle

\author
S.R.~Beane
Department of Physics, University of Maryland 
College Park, MD 20742-4111
{\it sbeane@physics.umd.edu}
\endauthor

\abstract
\singlespaced
\widenspacing

I show that in the large-$N_c$ limit the ground state baryons with
helicity $\lambda$ fall into an $(({N_c}- 2\lambda )/4,({N_c}+
2\lambda)/4)$ irreducible representation of $SU(2)\times SU(2)$.  This
representation determines the absolute normalization of the ground
state baryon axial vector couplings at large-$N_c$.  Results map
precisely to (spin-flavor) $SU(4)$ results.  For instance, I find
$\ga =(\nc +2)/3$.  As a consequence of this multiplet structure,
chiral symmetry forbids pion transitions between the ground state
baryons and other baryon towers in the large-$N_c$ limit.

\endabstract 
\vskip0.5in
\center{PACS: 11.30.Rd; 12.38.Aw; 11.15.Pg} 
\endtitlepage
\vfill\eject                                     % new page
% introduction
\superrefsfalse
\singlespaced
\widenspacing

With $N_c$ large but finite, the ground state baryons are arranged in
a degenerate tower with $I=J=\oneht,\threeht,\ldots
,\nct$\ref{witten}.  The axial vector couplings of these baryons grow
with $\nc$ and therefore that these baryons not violate unitarity
bounds in their interactions with pions implies a set of consistency
conditions\ref{gervais}. These consistency conditions have an
algebraic structure which maps to a contracted $SU(4)$ spin-flavor
symmetry. This symmetry determines {\it inter alia} ratios of
axial vector couplings.  Perhaps most importantly, in this algebraic
approach systematic $1/\nc$ corrections are straightforward to
incorporate\ref{dashen}.

On the other hand, the famed Adler-Weisberger (AW) relation is an
independent asymptotic constraint on pion-baryon
interactions\ref{adler}.  This constraint is more severe than that
implied by unitarity (when $N_c$ is large) and therefore it is
tempting to relegate the AW relation to the shadow world of model
dependence. However, the AW relation can also be expressed
algebraically. It corresponds to the statement that baryons fill out
representations of the full $SU(2)\times SU(2)$ algebra for each
helicity\ref{alg}. Hence what in the language of dispersion relations
appears to be a mysterious asymptotic constraint is in fact a
consequence of a symmetry of QCD (for a large-$\nc$ analysis of the AW
relation in the dispersion relation language, see \Ref{bron}). I will
simply incorporate the AW relation into the definition of spontaneous
symmetry breaking as follows. If $G$ is not a symmetry of the vacuum
but a subgroup $H$ is, then $G$ is said to be spontaneously broken to
$H$ and physical states fall into irreducible representations of $H$
{\it and} reducible or irreducible representations of $G$, {\it for
each helicity}.

In this note I will find the $SU(2)\times SU(2)$ representations
filled out by the ground state baryons in the large-$N_c$ limit. At
first glance this might appear over constraining.  After all, the
contracted $SU(4)$ spin-flavor symmetry determines ratios of ground
state baryon axial vector couplings. I will show that the $SU(2)\times
SU(2)$ representations lead to predictions for the ground state
baryons which are consistent with the contracted $SU(4)$ spin-flavor
symmetry and identical to those implied by the full uncontracted
spin-flavor $SU(4)$. Chiral symmetry therefore fixes the overall
normalization of the axial vector couplings with no input from the
quark model.  Of course the chiral predictions must be consistent with
those of contracted $SU(4)$ since $SU(2)\times SU(2)$ is a symmetry of
large-$N_c$ QCD. My approach in this note differs from that of
\Ref{strong}, where it was shown that $SU(4)\times O(3)$ results of
the quark model can be reproduced using only basic information about
the large-$N_c$ approximation together with chiral symmetry {\it and}
special assumptions about the transformation properties of the baryon
mass matrix. This generates a solution which relates the various
helicities.  Here I adopt the large-$N_c$ approximation and chiral
symmetry and determine the chiral representations for each helicity.

For the purposes of this note the underlying perturbative theory is
QCD with two massless flavors and $\nc$ colors which will ultimately
be taken large. The lagrangian has an exact $SU(2)\times SU(2)$ global
symmetry which is spontaneously broken to $SU(2)$ isospin (confinement
implies chiral symmetry breaking in the large-$\nc$
limit\ref{coleman}). The massless flavors of the underlying QCD
lagrangian transform as $(\oneht, 0)$ and $(0, \oneht )$ with respect
to $SU(2)\times SU(2)$.  How then do the baryons transform for each
helicity?  With $\nc =3$ the baryons can transform as combinations of
any number of $(\oneht, 0)$, $(0, \oneht )$, $(\threeht, 0)$,
$(0,\threeht )$, $(1,\oneht)$ and $(\oneht ,1)$ irreducible
representations.  With $\nc =5$, $(\fiveht, 0)$, $(0,\fiveht )$,
$(\oneht, 2)$, $(2,\oneht )$, $(1,\threeht)$ and $(\threeht ,1)$ are
also allowed representations. Generally, for a baryon made of $\nc$
(odd) quarks, there are $(\nc +1)(\nc +3)/4$ possible irreducible
representations.

I will take as a working assumption that for each helicity the ground
state baryons fill out a complete irreducible representation.  For
each helicity $\lambda$ the ground state baryon tower contains
isospins $I=|\lambda |, |\lambda |+1,\ldots ,\nct$.  The
representation with the lowest helicity, say $\lambda =\oneht$,
contains each isospin in the tower only once.  For each $\nc$ there is
a single irreducible representation in which each isospin in the tower
appears only once. For $\nc =3$ this is $(\oneht ,1)$ since this
irreducible representation contains isospins $I=\oneht\otimes
1=\oneht\oplus\threeht$ which correspond to $N$ and $\Delta$.  For
$\nc =5$ this is $(1,\threeht )$ with isospins $I=\threeht\otimes
1=\oneht\oplus\threeht\oplus\fiveht$.  For general $\nc$ this
representation is given by $(({N_c}-1)/4,({N_c}+1)/4)$ with isospins
$I=({N_c}-1)/4\otimes ({N_c}+1)/4
=\oneht\oplus\threeht\oplus\fiveht\oplus\ldots\oplus{\nct}$.  The
$\lambda =-\oneht$ representation is then given by the parity
conjugate representation $(({N_c}+1)/4,({N_c}-1)/4)$.

The same argument applies to other helicities. As the helicity is
incremented by unity the number of isospins in the representation is
decreased by unity, beginning with the lowest isospin.  Hence with
$\nc =3$, the $\lambda =\threeht$ $\Delta$ is in the $(0,\threeht )$
representation which contains isospin $I=\threeht$ and the $\lambda
=-\threeht$ $\Delta$ is in the $(\threeht ,0 )$ representation.  With
$\nc =5$, the $\lambda =\threeht$ baryons are in the $(\oneht, 2)$
representation which contains isospins $I=\oneht\otimes
2=\threeht\oplus\fiveht$ and the $\lambda =\fiveht$ baryon is in the
$(0,\fiveht )$ representation which contains isospin $I=\fiveht$, etc.
One can easily convince oneself that, for each helicity $\lambda$, the
ground state baryons fall into an $(({N_c}- 2\lambda )/4,({N_c}+
2\lambda)/4)$ irreducible representation of $SU(2)\times SU(2)$.

The consequences of this representation can be found by taking matrix
elements of the $SU(2)\times SU(2)$ algebra and using the
Wigner-Eckart theorem to express the algebra as a set of equations for
reduced matrix elements\ref{alg}. Each irreducible representation has
a unique solution.  With $SU(2)\times SU(2)$ generators ${{\cal
Q}^5_a}$ and ${T_a}$, I define the axial vector coupling

\offparens
$$
\bra{\beta , \lambda}{{\cal Q}^5_a}\ket{\alpha , \lambda}=
[{X_a^\lambda}]_{\sss\beta\alpha}
\EQN xdefined$$\autoparens
where $\ket{\alpha , \lambda}$ is a physical hadron state of definite
helicity $\lambda$. Taking matrix elements of the $SU(2)\times SU(2)$
algebra and inserting a complete set of states then gives

\offparens
$$
[{X_a^\lambda},{X_b^\lambda}]_{\sss\beta\alpha}=
i{\epsilon_{abc}}[{T_c}]_{\sss\beta\alpha}.
\EQN aw$$\autoparens
This is a generalized AW relation. I now use the Wigner-Eckart
theorem to write

$$
\bra{{I_\beta},{m_\beta}}{X_{(m)}^\lambda}\ket{{I_\alpha},{m_\alpha}}=
{{C}_{{I_\alpha}1}}({I_\beta},{m_\beta};{m_\alpha},{m})
X^\lambda ({I_\alpha},{I_\beta})
\EQN we$$
where the $C$'s are Clebsch-Gordan coefficients and the $X^\lambda
({I_\alpha},{I_\beta})$ are reduced matrix elements.  Taking matrix
elements of \Eq{aw} between states of definite isospin and inserting a
complete set of states then gives a set of coupled equations for the
reduced matrix elements\ref{alg}.

Below I will focus on the $\lambda =\pm\oneht$ representation,
$(({N_c}\mp 1)/4,({N_c}\pm 1)/4)$. This representation contains each
isospin once and therefore leads to $\nc$ coupled equations for
$\nc$ unknowns.  In what follows I will drop the helicity index.
With $\nc =1$ there is a single equation with solution
$|{X({\oneht},{\oneht})}|=\sqrt{3/4}$.  With $\nc =3$ there are three
equations,

$$\EQNalign{
&4{X({\oneht},{\oneht})^2}-4{X({\threeht},{\oneht})^2}=3 
\EQN nthreeeq;a \cr 
&4{X({\threeht},{\threeht})^2}
+10{X({\threeht},{\oneht})^2}=15 
\EQN nthreeeq;b \cr 
&{X({\oneht},{\oneht})}
-\sqrt{5}{X({\threeht},{\threeht})}=0,
\EQN nthreeeq;c \cr} 
$$\autoparens 
with solution $|{X({\oneht},{\oneht})}|=\sqrt{25/12}$,
$|{X({\threeht},{\threeht})}|=\sqrt{5/12}$ and
$|{X({\threeht},{\oneht})}|=\sqrt{4/3}$.
With $\nc =5$ there are five equations,

$$\EQNalign{
&4{X({\oneht},{\oneht})^2}-4{X({\threeht},{\oneht})^2}=3 
\EQN nfiveeq;a \cr 
&4{X({\threeht},{\threeht})^2}
+10{X({\threeht},{\oneht})^2}
-9{X({\fiveht},{\threeht})^2}
=15 
\EQN nfiveeq;b \cr 
&14{X({\fiveht},{\threeht})^2}
+4{X({\fiveht},{\fiveht})^2}
=35 
\EQN nfiveeq;c \cr
&{X({\oneht},{\oneht})}
-\sqrt{5}{X({\threeht},{\threeht})}=0
\EQN nfiveeq;d \cr
&
\sqrt{3}{X({\threeht},{\threeht})}
-\sqrt{7}{X({\fiveht},{\fiveht})}=0,
\EQN nfiveeq;e \cr} 
$$
with solution $|{X({\oneht},{\oneht})}|=\sqrt{49/12}$,
$|{X({\threeht},{\threeht})}|=\sqrt{49/60}$,
$|{X({\fiveht},{\fiveht})}|=\sqrt{7/20}$,
$|{X({\threeht},{\oneht})}|=\sqrt{10/3}$ and
$|{X({\fiveht},{\threeht})}|=\sqrt{12/5}$.  It is easy to convince
oneself that the general solution for $I=\oneht ,\threeht$ transitions
with $\lambda=\pm\oneht$ is:

$$\EQNalign{
&|{X}^{\sss{\pm{1\over 2}}}({\oneht},{\oneht})|=
{{(\nc +2)}/{\sqrt{12}}}
\EQN gensol;a \cr 
&|{X}^{\sss{\pm{1\over 2}}}({\threeht},{\threeht})|=
{{(\nc +2)}/{\sqrt{60}}}
\EQN gensol;b \cr
&|{X}^{\sss{\pm{1\over 2}}}({\threeht},{\oneht})|=
\sqrt{{(\nc +5)(\nc -1)}/12}. 
\EQN gensol;c \cr} 
$$
All transitions for all helicities can be obtained in this way.  It is
straightforward to find the chiral lagrangian parameters

$$\EQNalign{
&\ga=
\sqrt{2}|\bra{\, p\,}\,{{X_{\sss{(+)}}^{\sss{\pm{1\over 2}}}}}\,\ket{\, n\,}|=
\sqrt{\textstyle{4\over 3}}|{X^{\sss{\pm{1\over 2}}}({\oneht},{\oneht})}|
={{{(\nc +2)}/3}}
\EQN onega;a \cr
&{\cal C}=
\sqrt{3}|\bra{\,\Delta^{++}\,}\,{{X_{\sss{(+)}}^{\sss{\pm{1\over 2}}}}}
\,\ket{\, p\,}|=
\sqrt{\textstyle{3}}|{X^{\sss{\pm{1\over 2}}}({\threeht},{\oneht})}|
=\sqrt{{(\nc +5)(\nc -1)}}/2.
\EQN onega;b \cr}
$$
Note that chiral symmetry fixes the normalization of the axial vector
couplings, up to a conventional phase.  These results are equivalent
to placing the ground state baryon tower in an irreducible
representation of $SU(4)$\ref{mano}\ref{alfaro}. In the large-$\nc$
limit $\ga$ and ${\cal C}$ satisfy identical Goldberger-Trieman
relations with $g_{\sss\pi NN}$ interchanged with $g_{\sss\pi
N\Delta}$$^1$\vfootnote1{In the chiral lagrangian of \Ref{hemmert},
$g_{\sss\pi N\Delta}$ represents the axial vector coupling and is
related to $\cal C$ by $g_{\sss\pi N\Delta}={\cal C}/\sqrt{2}$.}.  It
follows that

$$
{{g_{\sss\pi NN}}\over{g_{\sss\pi N\Delta}}}=
{{2|X({\oneht},{\oneht})|}\over{3|X({\threeht},{\oneht})|}}
={2\over 3}+O({1\over{\nc^2}}),
\EQN contrtest$$
as expected on the basis of the large-$\nc$ consistency
conditions\ref{mano}.  Consistency with the contracted $SU(4)$ results
justifies {\it a posteriori} the working assumption that the ground
state baryons are in complete irreducible representations of
$SU(2)\times SU(2)$.

Chiral symmetry gives an important result beyond fixing the
normalization of the axial vector couplings at large-$N_c$.  Only
states within a given $SU(2)\times SU(2)$ representation communicate
by pion emission and absorption\ref{alg}. This is the essential
content of the AW relation.  Since the ground state baryons fall into
complete $SU(2)\times SU(2)$ irreducible representations at
large-$\nc$, {\it there are no pion transitions between the ground
state baryons and other baryons towers}.  Put another way, the AW
relations for pion scattering on the ground state baryons are
completely saturated by states within the ground state multiplet with
no transitions to states outside the multiplet.  Therefore, for
instance, $g_{\sss\pi N{B^*}}=0$ where ${B^*}$ is any baryon not in
the ground state tower. This is an exact consequence of chiral
symmetry in the large-$N_c$ limit. By contrast, adopting large-$N_c$
alone, the excited baryon axial vector couplings to the ground state
are down by (fractional) powers of $N_c$, but are generally
non vanishing\ref{bron}\ref{pirjol}.

What then occurs away from the large-$N_c$ limit where $SU(4)$ is
broken but $SU(2)\times SU(2)$ remains unbroken for each helicity?
Consider the $N-\Delta$ mass splitting, which is a $1/{N_c}$ effect.
This mass splitting is not consistent with $N$ and $\Delta$ placed in
a single irreducible representation of $SU(2)\times
SU(2)$\ref{alg}\ref{alfaro}.  This means that the $SU(2)\times SU(2)$
representation containing $N$ and $\Delta$ changes as a function of
$N_c$. In particular, away from the large-$N_c$ limit this
representation must be reducible in order to allow mass splitting
within the multiplet and so involves at least one unknown mixing
angle\ref{alg}\ref{alfaro}\ref{beane}. Why this is so is a
mystery. But it does point at profound qualitative differences between
${N_c}=3$ QCD and large-$N_c$ QCD.

\vskip0.1in

%\vfill\eject % new page 

\noindent This work was supported by the U.S. Department of Energy grant
DE-FG02-93ER-40762. I thank C.K.~Chow, T.~Cohen, B.~Holstein and
M.~Luty for valuable conversations.

%\vfill\eject % new page 
\nosechead{References}% % no section number
%\addTOC{1}{References}{\folio}% % add to contents
%\global\def\HeadText{{\tenit References}}% % running head text
\ListReferences \vfill\supereject \end